# High-resolution Resonance Spin-flip Raman Spectroscopy of Pairs of Manganese Ions in CdTe

R.V. Cherbunin,[1,2] V.M. Litviak,[1] I.I. Ryzhov,[1,2] A.V. Koudinov,[1,3,*] S. Elsässer,[2] A. Knapp,[2] T. Kiessling,[2] J. Geurts,[2] S. Chusnutdinow,[4] T. Wojtowicz,[5] G. Karczewski[4]

[1]Spin Optics Laboratory, St.-Petersburg State University, 198504 St.-Petersburg, Russia

[2]Physikalisches Institut (EP3), Universität Würzburg, 97074 Würzburg, Germany

[3]Ioffe Institute, 194021 St.-Petersburg, Russia

[4]Institute of Physics, Polish Academy of Sciences, 02668 Warsaw, Poland

[5]International Research Centre MagTop, Institute of Physics, Polish Academy of Sciences, 02668 Warsaw, Poland

*We report the observation of tens of minor lines of the combinational spin-flip Raman scattering in a CdTe:Mn quantum well by means of the high-resolution optical spectroscopy. Classification of this manifold leads to four characteristic values of energy, that correspond to four different types of pair clusters of Mn ions: the nearest, second, third etc. neighbors. All the four energies show up in a single experiment with a very high precision, providing experimental grounds for a deeper understanding of the d-d exchange interactions in a diluted magnetic semiconductor and demonstrating the capacity of the employed method. The major (nearest-neighbor) exchange constant $J_1 = 6.15 \pm 0.05$K was found to consent with its previously reported value. Other detected characteristic energies are as follows ($\pm 0.05$K): $J_{(2)} = 1.80$K, $J_{(3)} = 1.39$K, $J_{(4)} = 0.81$K.*

Diluted magnetic semiconductors (DMSs) exhibit a pronounced response to external magnetic fields and a variety of other spin-related phenomena.[1,2] While materials belonging to this family retain practical prospects of several kinds, they are especially important as a testing area for magnetic interactions in crystals. Since the early era of the solid state physics,[3,4] there has been an interest and demand in the profound understanding of the exchange interactions of magnetic ions in insulators and semiconductors. In the context of classical DMSs, that is, mainly A2B6 crystals with a substitutional Mn, the analyses revealed superexchange as a dominant mechanism in 1980s, while possible contributions from a Bloembergen-Rowland exchange term have been remarked.[5,6,7] The rise of spintronics and new practical magnetic materials like (Ga,Mn)As led to a renewed interest in the fundamentals of the indirect exchange interaction. The last years saw the appearance of new generalizations and new approaches in the field.[8,9,10] Verification and comparison of these sophisticated theoretical concepts suffers from a shortage of experimental material. For the *d-d* exchange, in practice, the only value available more or less reliably from measurements is the value (and sign) of the nearest-neighbor exchange constant, $J_1$. Theoretical studies, however, need a more extensive experimental foundation.

---

[*] koudinov@orient.ioffe.ru

(Cd,Mn)Te pseudobinary solid solutions are model representatives of the family and perhaps the best-studied wide-gap DMSs. At any composition, they crystallize in a zinc blende lattice and show random occupation of the cation sites by substitutional $Mn^{2+}$ ions.[11] Each Mn ion bears a local spin magnetic moment (due to the internal $3d^5$ electrons) which, in addition, possesses small crystal-field anisotropy. So the local Mn spins are easily involved in various kinds of interactions, e.g., with the *s*- and *p*-band charge carriers or with each other.

In very dilute compositions (low manganese concentration $x \leq 0.1\%$), the vast majority of Mn substitutional ions sit well away from each other, leading, at a local scale, to a single-spin behavior and, in aggregate, to veritable paramagnetic properties. However, as $x$ increases, inter-manganese short-range interactions become important very soon, because a rapidly increasing fraction of Mn spins find themselves close to another Mn spin.[12] These interactions, due to their antiferromagnetic origin, impair the mean magnetic susceptibility and modify the magnetic-saturation behavior by the formation of a net of interacting spins.[13] Thus the Mn-to-Mn interactions are of major importance for the physics of the DMSs.

Because the Mn ions occupy regular cation sites of the matrix crystal, their neighborhood can be described in terms of discrete configurations, or types. The nearest-neighbor (NN) configuration is characterized by the largest value of the Mn-to-Mn exchange constant, which was quantified by several techniques. Early estimates were based on the Curie-Weiss temperature.[12,14] More reliable values follow from neutron-scattering data[15] and, especially, the magnetization-steps (MST) method,[16] in which the total magnetization of a sample in an increasing external magnetic field experiences a step-up every time the field becomes strong enough to increase the total spin of an antiferromagnetically coupled pair of Mn ions by another unity. While different approaches more or less agree in the value of the NN exchange constant $J_1$, the next-nearest-neighbor (NNN) interactions are studied not so reliably.[7,15,17,18,19,20] The reason is that the mentioned methods are *not really selective* in respect with the "sort of pairing" of Mn ions (i.e., the question whether the two involved Mn ions see each other in the first coordination sphere (NN locations), or the second, etc). As a result, the signals are usually dominated by contributions from single Mn's with an addition from the NN pairs. Existing estimates of the NNN exchange constants $J_2$, $J_3$, etc. are based on marginal discrepancies between the measured and calculated MST curves.

In the present communication, we report a selective approach to this problem. We recently demonstrated that the spectral response of resonant Raman scattering in DMS layers is much more complex than has been previously realized.[21] Here, we identify the nature of this complex response, i. e., selectivity of the scattering process with nearest neighbor ions, next-nearest neighbors and so on. We obtained up to 4 different characteristic energies in one measurement and with a remarkable precision of ±0.01 meV. This is a striking result in two regards: First, it shows that the accepted picture of resonant Raman scattering processes itself is at least incomplete. Second, and maybe even more important, we open a new door for a direct experimental effort towards distinguishing different magnetic coupling mechanisms.

Figure 1 shows a spectral map of the radiation from the QW in the coordinate frame 'energy – magnetic field', where the emission intensity is encoded by the points' color. The incident laser energy (1762.6 meV for Fig.1) is somewhat above the exciton luminescence band (maximum ~ 1755 meV, FWHM ~ 5 meV, visible as a blurry ocher yellow area). The sharp brown lines depict replicas of the resonance Rayleigh (the horizontal bar) and spin-flip Raman (inclined traces) scattering of the laser light. Note that all replicas are greatly enhanced in the vicinity of the exciton state, meaning that the mechanisms behind the scattering processes have a direct relationship to the QW exciton.[22,23,24] In Fig.1, the *n*-th inclined trace (counting from the Rayleigh line) corresponds to the light scattering process in which the scattered photon leaves behind *n* flipped spins of $3d^5$ electrons of the Mn ions. The inclination of the traces with regard to the magnetic field axis represents the magnetic-field dependence of the change of the scattered photon energy. This directly reflects the energy corresponding to the spin flip in the external magnetic field.[25]

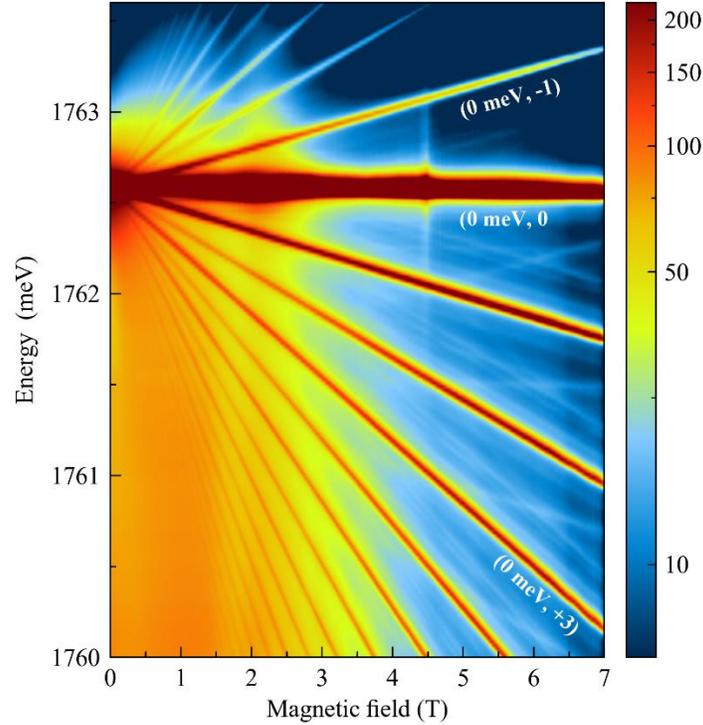

Figure 1. Spectral map of the radiation from the 9ML (3 nm) QW for *B*-fields up to 7 T, T=1.5 K, the laser excitation energy (1762.6 meV) hits the high-energy side of the QW exciton band (blurry ocher yellow area). The thick horizontal bar corresponds to the elastic light scattering, the sharp inclined lines below (above) that – to Stokes (anti-Stokes) multiple inelastic light scattering by paramagnetic spins of single Mn ions. A manifold of weak traces between the major lines represents the subject of the present study.

The fan-like system of traces in Fig.1 makes the typical picture of the multiple Raman paramagnetic resonance, alike observed many times in similar systems.[26,27,28,29,30,31] Almost 20 traces below the laser energy correspond to Stokes processes (those in which the energy of the crystal is increased), a dozen of traces above the laser – to the anti-Stokes processes.

However, in marked contrast to the previous studies, the high-resolution experiment reveals tens of minor traces which do not belong to the main fan. These traces are visible in some areas of the map and disappear in other areas; some of them make scissor-like structures, some appear as twinned lines through a long path, or manifest themselves as close satellites of the strong lines of the main fan (Fig.2 (a)). While some elements clearly made repeating patterns, they did not secure an immediate perception of the entire picture. Ultimately, we found a scheme to classify the variety of minor traces by affiliating them to several additional fan-like features. Each of the fans roots at a unique point in energy (above or below the laser) at zero magnetic field, and each one includes a significant number of traces (Fig.2 (b) – (d)). Each root energy represents the spin-flip-induced energy difference of a spin system in its zero-field extrapolation, i.e., after correction for the Zeeman term. Therefore it is attributed to the exchange interaction.

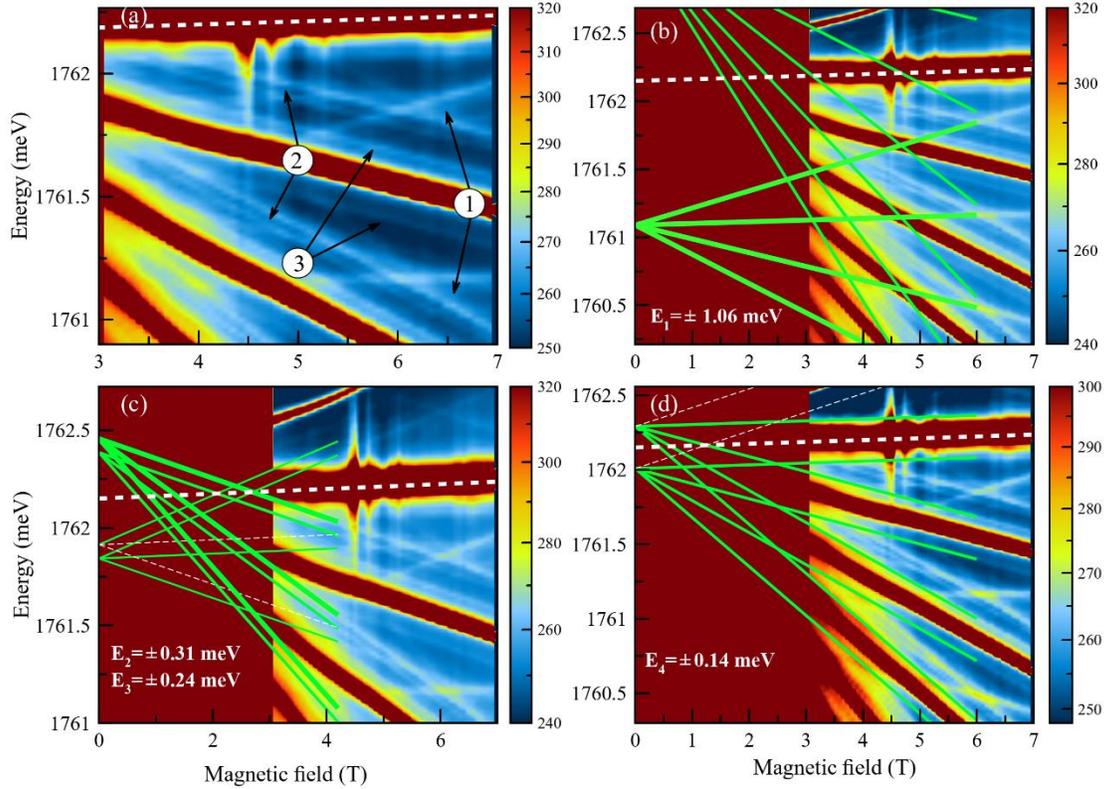

Figure 2. Resonance Raman scattering in a (Cd,Mn)Te QW: a detailed presentation of some regions of the experimental intensity maps (similar to Fig.1). Panel (a) illustrates typical repeating patterns of the manifold: 1 – "scissors", 2 – twinned traces, 3 – satellite weak lines. Panels (b) to (d) depict the fitting of the key

elements of the observed manifold by additional fan-like structures. Thick dotted line in every panel marks the exact position of the peak of the excitation laser.

Let us introduce a notation for specific fan-making traces. Let the trace ($E$, $n$) be characterized by the energy $E$ of the fan root (i.e., at $B = 0$) and by an integer $n$ indicating the slope toward the $B$-field axis. With this notation, the major fan in Fig.1 will be designated as (0 meV, $n$) and will include the Rayleigh line (0 meV, 0) and, for example, the third Stokes Raman line (0 meV, +3) and the first anti-Stokes Raman component (0 meV, -1), see Fig.1.

Inspection of minor traces shows that many of them belong either to the fan (+1.06 meV, $n$) or to its anti-Stokes partner (-1.06 meV, $n$), see Fig. 2 (b). Each family includes about 10 members – we note, positively more than three. The experimental traces (+1.06 meV, $n$) are particularly well-seen; for instance, they form upper elements of the "scissors" at high fields. The traces (-1.06 meV, $n$) are less pronounced, even when their total Raman shifts are about those of the corresponding (+1.06 meV, $m$) traces. This can be ascribed to the fact that the processes (-1.06 meV, $n$) probably include absorption of a 1.06-meV excitation within the crystal; that is, the resulting Raman processes are not fully independent of energy in other reservoirs (such as the lattice or Mn systems) but require an available portion of energy, missing at helium temperatures.

Other very pronounced elements of the observed manifold of traces belong to the fans (-0.31 meV, $n$) and (-0.24 meV, $n$). The former of them make the lower elements of the "scissors", while in combination, they form the long twinned trace in the "scissors handle" (see Fig.2 (c)). Again, the number of traces in both fans is definitely more than three. Strange enough, the Stokes partners of the above fans are barely visible. One can just notice traces (+0.31 meV, $n$) with $n = 0, \pm 1$, while the trace (+0.24 meV, -1) is only visible above the Rayleigh line. Traces (-0.31 meV, 0) and (-0.24 meV, 0) are not visible, for some reason.

Finally, the remaining traces can be attributed to the fan (+0.14 meV, $n$) and its anti-Stokes partner (-0.14 meV, $n$). They appear as close satellites on either side of the intense (0 meV, $n$) traces (see Fig 2 (d)). For both the Stokes and anti-Stokes fans, only traces with positive numbers were detectable.

Thereby, the experimental traces point, for vanishing $B$-field, to four well-defined values of energy in the sub-meV range which are characteristic for the system under study. The only candidates in sight to provide such energies are pairs of coupled Mn ions.[32,33,34,35]

In order to compare our values of the exchange integrals with the data of previous authors, let us assume each observed energy equal to a doubled exchange constant (of the corresponding sort), and we will obey the tradition of expressing their values in Kelvins (1 K ~ 0.086 meV). In such a way, we report $J_{(1)} = 6.15K, J_{(2)} = 1.80K, J_{(3)} = 1.39K, J_{(4)} = 0.81K$. (Here, as usual, parenthesized indices are used to designate exchange constants which are ranged only by their descending value, not by the increasing distance between the ions constituting the pair.) Basically, this agrees with the previous knowledge regarding various DMSs and, in particular, (Cd,Mn)Te.[36] The largest measured quantity $J_1 = J_{(1)} = 6.15K$ is in excellent agreement with the previously reported value of the F=0 to F=1 excitation of a pair of NN manganese spins, $J_1 \approx 6.1K$ or $J_1 = 6.2K$.[37,38,39] Both the values $J_{(2)}$ and $J_{(3)}$ hit the interval of expectations (1-2 K) for the NNN constants;[17-18,36] the same refers to $J_{(4)}$. Excellent precision (about $\pm 0.05$ K) is an obvious advantage of the Raman method. But the latter does not allow for an easy attribution of the observed exchange constants to the specific types of pair clusters. We suggest a combined study, in which the Raman-obtained *values* would be a prerequisite; then different *attributions* of $J_{(2)}, J_{(3)}, J_{(4)}$ to $J_2, J_3, J_4$ would represent models for the simulation of the MST curves (in the spirit of Refs.[17,19,20]); finally, one of the resulting MST curves would be preferred by comparison with the experimental MST data. We haven't perform a cluster calculation for a strange-sounding reason: we did not find in the literature a suitable magnetic-field dependence of (Cd,Mn)Te magnetization to compare the calculation with! Indeed, the specially-measured field dependence of Ref.[20], which would be ideal in many respects (small step in $B$, low temperature, low Mn concentration), unfortunately covers only the fields' range 0-5 T, while for our deduced values of the NNN exchange constants, the three pentads of magnetization steps are expected to populate a much wider range 0-17 T. In particular, within the 0-5 T range, one can expect to see as few as zero steps due to the $J_1$ and no more than one step due to deduced $J_{(2)}$ and $J_{(3)}$. The little amplitude of a single step, its large thermal broadening and the absence of observable periodicity in *B* makes the steps hardly distinguishable. Therefore, observation of several steps is essential, thus calling for a perfect magnetization measurement in a range 0-20 T. However, for this range of the fields, numerous published magnetization measurements are not precise enough and, often, use indirect (e.g., optical) approach to the quantification of the magnetization. Such excitonic measurements can easily be criticized when so very small effects in the magnetization are in focus. Nevertheless, we note that by analyzing the optically-characterized magnetization (in the proper fields' range 0-20 T) by means of a cluster approach, the authors of Ref.[17] came to the values $|J_1| = 6.3$ K (cf. our $J_1 = 6.15$ K) and $|J_2| = 1.9$ K (cf. our $J_{(2)} = 1.80$ K). Their third reported constant, $|J_3| = 0.4$ K, does not match our $J_{(3)}$ or $J_{(4)}$ so good.

In summary, by the application of the high-resolution resonance multiple $Mn^{2+}$ spin-flip Raman spectroscopy we observed tens of never-reported Raman lines in the spectra of the model DMS QWs in the conditions of the Raman paramagnetic resonance. Magnetic-field dependence of the positions of the lines revealed four different characteristic energies of the system. We attribute these energies to pair clusters of $Mn^{2+}$ ions whose spins are coupled by an exchange interaction. Thus we report four exchange constants of the pair interaction in a CdTe:Mn quantum well with a remarkable experimental precision. Overall, the straightforward interpretation of the experimental Raman traces looks satisfactory; a little cloud on the horizon is the unclearly low intensity of the (+0.31 meV, *n*) and (+0.24 meV, *n*) fans with respect to their anti-Stokes partners.

The present study solves the previous puzzle of the "fractional peaks" which was reported in Ref.[21]. Obviously, the "half-integer" positions of the minor Raman lines in Ref.[21] originated from the coincidental crossing of (+1.06 meV, *n*) and (-0.31 meV, *n* + 3) traces right between the (0 meV, $n + 1$) and (0 meV, $n + 2$) traces of the major fan (at the field $B \approx 6$ T). In addition to that, the "half-integer" slopes between the traces ($n + 1$) and ($n + 2$) were imitated by an average of the slopes $n$ and ($n + 3$) of the unresolved minor peaks.

We note that the experimental technique reported here can become a prototype for the accurate quantitative investigation of the exchange interactions between the neighboring spins in various crystals. The Raman paramagnetic resonance effect was already reported for several semiconductor systems (i.a., not only with substituting Mn)[40,41]. Essential elements for the method would be: a narrow QW showing an exciton resonance, an appropriate concentration of magnetic ions, a magnetic field and a high-resolution Raman-scattering experiment.

**Methods**

The samples for the present study were conventional MBE-grown nanostructures of a common design (as described, e.g., in Ref.[21]). They are all grown on GaAs substrates followed by a several (3 – 4) micron thick CdTe buffer. In essence, they included narrow (3 or 4.5 nm) isolated (Cd,Mn)Te quantum wells (QWs) with symmetric non-magnetic barriers of (Cd,Mg)Te containing ~20% of Mg. The Mn concentrations inside the QW layers were different but did not exceed few percent. We chose for demonstration the experimental results obtained using the sample A described earlier (1.7% of Mn).[21]

We performed an experimental multiple $Mn^{2+}$ paramagnetic resonance Raman study using the original configuration, as described in Ref.[21]. Namely, the external magnetic field was parallel to the plain of the QWs while the light emission was collected in the backscattering geometry. However, the resolution of the experiment was beyond standard: owing to the use of a narrow-band laser (Coherent MBR-110) and a triple grating spectrometer Dilor XY-500 in the additive mode, we achieved an overall spectral resolution of about 25 μeV (0.2 cm$^{-1}$), which is, e.g., ~ 6 times smaller than in Ref.[21]. We discovered that such instrumental improvement immensely enriches the spectral picture and the informative value of the effect. Unfortunately, in the express presentations of the above results, we reported slightly different values of the $J_1$-$J_4$.[42] These 20% lower values originated from a calibration error of the spectrometer. As mentioned, the correct value of $J_1$ is much more consistent with the previous studies.

**Acknowledgements**

This work was partially supported by a grant from Saint-Petersburg State University and DFG (Project No.40.65.62.2017), by DFG-project Ge 1855/14-1 and by RFBR (project 19-02-00422). IIR acknowledges support from RFBR project 19-52-12054. The research in Poland was partially supported by the Foundation for Polish Science through the IRA Programme co-financed by EU within SG OP and by the National Science Centre through Grant No. UMO 2017/25/B/ST3/02966.
We acknowledge useful notes by J. Kossut and V.F. Sapega.

**REFERENCES**

[1] *Introduction to the Physics of Diluted Magnetic Semiconductors*, ed. by J. Gaj and J. Kossut. Springer-Verlag Berlin Heidelberg 2010.

[2] *Diluted Magnetic Semiconductors*, edited by J. Furdyna and J. Kossut, *Semiconductors and Semimetals* Vol. 25, edited by R. K. Willardson and A. C. Beer, Academic, New York, 1988.

[3] H.A. Kramers. Physica 1, 182 (1934)

[4] P.W. Anderson. Phys. Rev. 79, 350 (1950)

[5] J. Spałek, A. Lewicki, Z. Tarnawski, J.K. Furdyna, Z. Obuszko. Phys. Rev. B 33, 3407 (1986)

[6] B.E. Larson, K.C. Hass, H. Ehrenreich, and A.E. Carlsson. PRB 37, 4137 (1988)

[7] A. Lewicki, J. Spałek, J. Furdyna, and R. Gałazka. Phys. Rev. B 37, 1860 (1988)

[8] A. Savoyant, S. D'Ambrosio, R.O. Kuzian, A.M. Daré, and A. Stepanov. Phys. Rev. B 90, 075205 (2014)

[9] T. Linneweber, J. Bünemann, U. Löw, F. Gebhard, and F. Anders. PRB 95, 045134 (2017)

[10] C. Śliwa and T. Dietl. PRB 98, 035105 (2018)

[11] Y. Shapira, S. Foner, D. H. Ridgley, K. Dwight, and A. Wold. Phys. Rev. B 30, 4021 (1984)

[12] S. Oseroff, P.H. Keesom, Magnetic properties: macroscopic studies, in Ref.[2], pp.73-123

[13] D. Heiman, E. D. Isaacs, P. Becla, and S. Foner. Phys. Rev. B 35, 3307 (1987)

[14] W. Y. Ching, D. L. Huber. Phys. Rev. B 30, 179 (1984)

[15] T. M. Giebultowicz, J. J. Rhyne, W. Y. Ching, D. L. Huber, J. K. Furdyna, B. Lebech, and R. R. Galazka. Phys. Rev. B 39, 6857 (1989)

[16] For review, see Y. Shapira, V. Bindilatti. J. Appl. Phys. **92**, 4155 (2002)

[17] B. E. Larson, K. C. Hass, and R. L. Aggarwal. Phys. Rev. B 33, 1789 (1986)

[18] Xiaomei Wang, D. Heiman, S. Foner, and P. Becla. Phys. Rev. B 41, 1135 (1990)

[19] V. Bindilatti, E. ter Haar, N. F. Oliveira, Jr., Y. Shapira, and M. T. Liu. Phys. Rev. Lett. 80, 5425 (1998)

[20] H Malarenko, V Bindilatti, N.F Oliveira, M.T Liu, Y Shapira, and L Puech. Physica B: Condensed Matter 284-288, 1523 (2000)

[21] A. V. Koudinov, A. Knapp, G. Karczewski and J. Geurts. Phys. Rev. B 96, 241303(R) (2017)

[22] J. Stühler, M. Hirsch, G. Schaack, and A. Waag. Phys. Rev. B 49, 7345 (1994)

[23] A. V. Koudinov, Yu. G. Kusrayev, B. P. Zakharchenya, D. Wolverson, J. J. Davies, T. Wojtowicz, G. Karczewski, and J. Kossut Phys. Rev. B 67, 115304 (2003)

[24] A. V. Koudinov, Yu. G. Kusrayev, D. Wolverson, L. C. Smith, J. J. Davies, G. Karczewski, and T. Wojtowicz. Phys. Rev. B 79, 241310(R) (2009)

[25] A. Petrou, D. L. Peterson, S. Venugopalan, R. R. Galazka, A. K. Ramdas, and S. Rodriguez. Phys. Rev. Lett. 48, 1036 (1982) and A. Petrou, D. L. Peterson, S. Venugopalan, R. R. Galazka, A. K. Ramdas, and S. Rodriguez. Phys. Rev. B 27, 3471 (1983); for review, see A.K. Ramdas, S. Rodriguez. Raman scattering in Diluted Magnetic Semiconductors. in Ref.[2], pp.345-412

[26] I. I. Reshina, S. V. Ivanov, D. N. Mirlin, A. A. Toropov, A. Waag, and G. Landwehr Phys. Rev. B 64 035303 (2001)

[27] M. Byszewski, D. Plantier, M.L. Sadowski, M. Potemski, A.Sachrajda, Z.Wilamowski, G.Karczewski. Physica E 22, 652 (2004)

[28] J. M. Bao, A. V. Bragas, J. K. Furdyna, and R. Merlin. Phys. Rev. B 71, 045314 (2005)

[29] L. C. Smith, J. J. Davies, D. Wolverson, M. Lentze, J. Geurts, T. Wojtowicz, and G. Karczewski. Phys. Rev. B 77, 115341 (2008)

[30] C. Kehl, G. V. Astakhov, K. V. Kavokin, Yu. G. Kusrayev, W. Ossau, G. Karczewski, T. Wojtowicz, and J. Geurts. Phys. Rev. B 80, 241203(R) (2009)

[31] N. V. Kozyrev, R. R. Akhmadullin, B. R. Namozov, Yu. G. Kusrayev, I. V. Sedova, S. V. Sorokin, and S. V. Ivanov. Phys. Rev. B 99, 035301 (2019)

[32] D.U. Bartholomew, E.-K.Suh, S.Rodriguez and A.K. Ramdas. Solid State Commun. 62, 235 (1987)

[33] J. Stühler, G. Schaack, M. Dahl, A. Waag, G. Landwehr, K. V. Kavokin and I. A. Merkulov. J. Raman Spectroscopy 27, 281 (1996)

[34] J Puls and F Henneberger. J. Cryst. Growth 214-215 432 (2000)

[35] Worth noting, we observed no minor traces in similar samples with lower Mn concentrations $x \cong 0.1\%$, while the major fan with many traces was visible quite clear. This is in agreement with a lower occurrence of isolated pairs, as compared to isolated single ions, in very dilute DMSs (R.E. Behringer. J. Chem. Phys. 29, 537 (1958)).

[36] J.A. Gaj, J.Kossut. Basic consequences of the *sp-d* and *d-d* interactions in DMS. In Ref.[1], see pp.31-34.

[37] Y. Shapira and N. F. Oliveira, Jr. Phys. Rev. B 35, 6888 (1987)

[38] S. Foner, Y. Shapira, D. Heiman, P. Becla, R. Kershaw, K. Dwight, and A. Wold Phys. Rev. B 39, 11793 (1989)

[39] A. I. Savchuk, V. I. Fediv, P. I. Nikitin, A. Perrone, O. M. Tatzenko, and V. V. Platonov. J. Cryst. Growth 184–185, 988 (1998)

[40] M.J. Seong, H. Alawadhi, I. Miotkowski, A.K. Ramdas, and S. Miotkowska. Phys. Rev. B 63, 125208 (2001)

[41] X. Lu, S. Tsoi, I. Miotkowski, S. Rodriguez, A. K. Ramdas, and H. Alawadhi Phys. Rev. B 75, 155206 (2007)

[42] https://www.jaszowiec.edu.pl/2019/abs/FrO4.pdf; https://arxiv.org/abs/1903.01276v1